\documentclass[twocolumn,showpacs,preprintnumbers,amsmath,amssymb]{revtex4-1}
\usepackage{graphicx}
\usepackage{dcolumn}
\usepackage{bm}

\begin{document}

\preprint{APS/123-QED}

\title{Time Resolved Measurement of Electron Cloud Densities 
from Dispersion of Transverse Electric Pulses}
\author{Kiran G. Sonnad}
\affiliation{High Energy Accelerator Research Organization (KEK),
Oho 1-1, Tsukuba, Ibaraki 305-0801, Japan}

\begin{abstract}
The measurement of electron cloud densities in particle accelerators using 
microwaves has proven to be an effective, non-invasive and inexpensive
method. So far the experimental schemes have used continuous waves. 
This has either been in the form of travelling waves that are propagated, 
or standing waves that are trapped, in both cases within a segment of 
the accelerator chamber. The variation in the wave dispersion relation 
caused by the periodic creation and decay of the electron cloud leads 
to a phase modulation in the former case, and a frequency modulation 
in the latter. In general, these methods enable the measurement of a 
time averaged electron cloud density.  In this paper we propose a time 
resolved measurement by using pulses propagated over a finite length of 
the accelerator chamber. The pulses are launched periodically, once after 
a bunch train has passed and then again half a revolution period later.  
This results in pulses alternating between a dispersion that is either
affected by a cloud or not. The resulting spectrum of the signal can
be related to the electron density sampled by the pulse that propagates
through the cloud. By varying the delay of the launch of the pulse with 
respect to the train passage, one can map the cloud density at different points
behind the train.    
\end{abstract}


\date{\today}

\maketitle

\section{Introduction}
A better understanding of electron clouds is of prime importance for
the optimal performance of a number of present and future accelerators.   
One area of this study includes methods of measuring the density 
of the cloud. The idea of using the dispersion of microwaves caused by    
electron clouds, acting like a plasma was first introduced by F. Caspers
\cite{Caspers2004Napa,kroyer2005cern}. Since then a number of related 
studies have been conducted based on this principle. Overall, these studies 
can be divided into two categories (1) The use of travelling
waves \cite{DeSantis-etal} and (2) the use of standing waves \cite{Sikora-etal}. 
Both the methods involve generating a continuous wave, and they typically 
provide an electron cloud density estimate that is integrated over time. 
In the former case, the electron cloud produces phase modulation of the wave 
while in the latter it is frequency modulation. The modulation occurs at the 
revolution frequency of the bunch train. In both cases, this leads to the 
production of sidebands that are spaced at harmonics of the revolution 
frequency off the carrier wave frequency. 

Instead of a continuous wave, this paper proposes studying the 
propagation of a periodic pulse, and shows how the resulting signal 
produced from such a process, can be utilized in determining the 
electron cloud density at various points behind a train of bunches 
in a storage ring. It assumes that there is a single train in the ring, 
and that all the cloud clears away rapidly enough at the point of 
interest, after the train passes. A pulse is propagated over a short 
distance every turn, and with a certain delay following the train 
passage. Another identical pulse is propagated at half a revolution 
period later, when the train is half way around in the ring. In 
principle, the resulting spectrum of these periodic pulses combined 
would obey the relationships provided in section IV. By changing the 
delay of the pulse, one can obtain information of the cloud density 
at different points behind the train. To establish the feasibility of 
performing such a time resolved measurement, we propose replicating 
such a process through a table-top experiment, representing a ``static" 
uniform cloud of constant density  that abruptly appears and disappears. 
In the next section we discuss the analysis of this much simpler
scenario. In section IV, necessary modifications in the analysis of section II 
are presented, when performing such an experiment in an operating storage 
ring where the cloud is ``dynamic", {\it ie} it varies with time and position. 

\section{Alternating pulses with a static perturbation in the dispersion}
Suppose we produce a pulse with shape $f(t)$ that has a periodicity $T_0$.
Then the signal produced by the series of pulses may be expressed as 
\begin{equation}
 F(t) = \sum_{n=-\infty}^{\infty}f(t - nT_0)
\end{equation} 
We may express $F(t)$ such that
\begin{equation}
 F(t) = \sum_{n=-\infty}^{\infty}F_n{\rm e}^{in\omega_0 t}
\end{equation}
where $\omega_0 = 2\pi/T_0$. We can then see that,
\begin{equation}
F_n = \frac{1}{T_0} \int^{T_0}_{0} dt F(t){\rm e}^{-in\omega_0 t}
\end{equation}
If we substitute the above expression for $F(t)$, we have
\begin{equation}
 F_n = \frac{1}{T_0} \int^{T_0}_{0}dt \sum^{\infty}_{m=-\infty}f(t-mT_0)
 {\rm e}^{-in\omega_0 t}
\end{equation}
If we substitute $t = t^{\prime} + mT_0$ and interchange the order of summation
and integration, we have
\begin{eqnarray}
 F_n &=& \frac{1}{T_0} \sum_{m=-\infty}^{\infty} \int_{-mT_0}^{-(m-1)T_0}
 dt^{\prime}f(t^{\prime}){\rm e}^{-in\omega_0t^{\prime}} \nonumber \\ 
  &=& \frac{1}{T_0}\int_{-\infty}^{\infty} 
 dt^{\prime}f(t^{\prime}){\rm e}^{-in\omega_0t^{\prime}} = \frac{1}{T_0}\tilde{f}(n\omega_0) 
\label{HarmAmplitude}
\end{eqnarray} 
where $\tilde{f}(\omega)$ is the Fourier transform of $f(t)$ given by
\begin{equation}
  \tilde{f}(\omega) = \int_{-\infty}^{\infty}dt f(t){\rm e}^{-i\omega t} 
\end{equation}
and consequently, the inverse transform is given by
\begin{equation}
  f(t) = \frac{1}{2\pi}\int_{-\infty}^{\infty} d\omega \tilde{f}(\omega) {\rm e}^{i\omega t}
\end{equation}

Suppose the generated pulse is allowed to travel through a dispersive medium, 
then the shape of the pulse would evolve as 
\begin{equation}
 f(t,x) = \frac{1}{2\pi} \int_{-\infty}^{\infty} d\omega \tilde{f}(\omega) {\rm e}^{-i(kx - \omega t)}  
\label{f-tx} 
\end{equation} 
where $k = k(\omega)$ is a function of $\omega$ and is given by the dispersion relationship of the medium. If the 
relationship between $k$ and $\omega$ is linear, we know that the shape of the pulse is preserved. We may express 
this propagated signal as
\begin{equation}
 F(t,x) = \sum^{\infty}_{n=-\infty}F_n(x){\rm e}^{in\omega_0t} 
\end{equation}
where $F_n(0) = F_n$. It is easy to show that $F_n(x) = F_n{\rm e}^{-ik(n\omega_0)x}$.  
Using Eq~(\ref{HarmAmplitude}), we have
\begin{equation}
 F_n(x) = \frac{1}{T_0}\tilde{f}(n\omega_0,x)
\end{equation}
Using Eq~(\ref{f-tx}) and the definition of Fourier transform, this gives
\begin{equation}
 F_n(x) = \frac{1}{2\pi T_0} \int_{-\infty}^{\infty} \int_{-\infty}^{\infty} dt d\omega \tilde{f}(\omega)
 {\rm e}^{-i(kx - \omega t)} {\rm e}^{-in\omega_0t}  
\end{equation}
Using the standard expression for the dirac-delta function $\delta(a-\alpha) = 
(1/2\pi)\int^{\infty}_{-\infty}dp{\rm e}^{ip(a-\alpha)}$
we get
\begin{eqnarray}
 F_n(x) &=& \frac{1}{T_0}\int_{-\infty}^{\infty} d\omega \tilde{f}(\omega) 
{\rm e}^{-ik(\omega)x}\delta(\omega - n\omega_0) \nonumber \\ 
  &=& \frac{1}{T_0}\tilde{f}(n\omega_0) {\rm e}^{-ik(n\omega_0)x} = F_n{\rm e}^{-ik(n\omega_0)x} 
\label{Snxexp}
\end{eqnarray}

We now study the case where we generate a pulse between intervals of $T_0/2$. The pulse is allowed to
propagate a distance $x$ before being detected. We further assume that the dispersion relationship varies slightly between 
alternate pulses. Assume that at one instance of pulse propagation the wave number is $k(\omega)$, and at 
the next it is $k(\omega) + \delta k(\omega)$, with this pattern repeating itself. We denote the modification
of the pulse shape due to $\delta k(\omega)$ as $f + \delta f$, and that of the corresponding amplitude of the $n$th 
harmonic as $F_n + \delta F_n$. The signal is then given by
\begin{equation}
 F(t,x,\delta k) =  \sum^{\infty}_{n=-\infty}[f(t-\frac{nT_o}{2},x) + \delta f(t-nT_0,x)]
\label{doubspect}
\end{equation}   
where $\delta f(t,0) = 0$. This leads us to
\begin{equation}
F(t,x,\delta k) = \sum^{\infty}_{n=-\infty} F_{2n}(x){\rm e}^{i2n\omega_0 t} + 
\sum^{\infty}_{n=-\infty} \delta F_n(x){\rm e}^{in\omega_0 t}
\label{doubspectFt}
\end{equation}
where we have in analogy to Eq~(\ref{HarmAmplitude}),
\begin{equation}
 F_{2n}(x) = \frac{2}{T_0}\tilde{f}(2n\omega_0,x)
\label{S2nexp}
\end{equation}
and
\begin{equation}
 \delta F_n(x) = \frac{1}{T_0}\tilde{\delta f}(n\omega_0,x)
\label{df-signal}
\end{equation}
Eq~(\ref{doubspectFt}) may be rewritten as
\begin{eqnarray}
  F(t,x,\delta k) & & = \sum^{\infty}_{n=-\infty}[F_{2n}(x) 
  + \delta F_{2n}(x)]{\rm e}^{i2n\omega_0 t} \nonumber \\ 
  &+&  \sum^{\infty}_{n=-\infty}
 \delta F_{2n-1}(x){\rm e}^{i(2n-1)\omega_0 t} 
\end{eqnarray}
Thus, at every even harmonic of $\omega_0$, we see a signal amplitude of $F_n(x) + \delta F_n(x)$ and at every
odd harmonic we see a signal amplitude of $\delta F_n(x)$. We proceed to derive a relationship between the
ratio of even and odd harmonic amplitudes, and the corresponding value of $\delta k$ at that harmonic.   
Analogous to Eq~(\ref{f-tx}), we have 
\begin{equation}
 f(x,t) + \delta f(x,t) = \frac{1}{2\pi}\int^{\infty}_{-\infty} d\omega \tilde{f}(\omega) 
{\rm e}^{-i[k(\omega)x+\delta k(\omega)x - \omega t]}   
\end{equation} 
A Taylor expansion to first order in $\delta k(\omega)x$ gives
\begin{eqnarray}
 & & f(x,t) + \delta f(x,t) \approx \frac{1}{2\pi}\int^{\infty}_{-\infty} d\omega \tilde{f}(\omega) 
{\rm e}^{-i[k(\omega)x - \omega t]} \nonumber \\ 
& & \times [1 - i\delta k(\omega)x] 
\end{eqnarray}
This gives to first order
\begin{equation}
  \delta f(x,t) = -\frac{ix}{2\pi}\int^{\infty}_{-\infty} d\omega \delta k(\omega) \tilde{f}(\omega) 
{\rm e}^{-i[k(\omega)x - \omega t]}
\label{df-expr}
\end{equation}
Taking the Fourier transform of this, and following the same process as before of using the expression for 
the dirac-delta function as in Eq~(\ref{Snxexp}) we 
get
\begin{eqnarray}
 \delta F_n(x) &=& -\frac{1}{T_0}ix\delta k(n\omega_0)\tilde{f}(n\omega_0){\rm e}^{-ik(n\omega_0)x} \nonumber \\ 
 &=&  -\frac{1}{T_0}ix\delta k(n\omega_0)\tilde{f}(n\omega_0,x)
\end{eqnarray} 
Using the above equation and Eq~(\ref{S2nexp}), we obtain the ratio between an even and an odd harmonic amplitude to be
\begin{eqnarray}
   R_{mn} &=& \frac{|\delta F_{2m-1}(x)|}{|F_{2n}(x) + \delta F_{2n}(x)|} \approx 
\frac{|\delta F_{2m-1}(x)|}{|F_{2n}(x)|}  \nonumber \\ 
&=& x\delta k([2m-1]\omega_0)\frac{|\tilde{f}[(2m-1)\omega_0,x]|}{2|\tilde{f}(2n\omega_0,x)|}  
\label{sig-ratio} 
\end{eqnarray}
Thus, $R_{mn}$ depends on the product of two terms. (1) The quantity $x\delta k([2m-1]\omega_0)$ which is 
the phase shift  induced by the variation in the dispersion 
relation for the wave at frequency $(2m-1)\omega_0$ after it propagates a distance $x$.
(2) The ratio between the Fourier transform of the shape of the unperturbed pulse evaluated at the two frequencies
corresponding to the even and odd harmonics respectively. The latter quantity can be computed from the shape
of the pulse, or it can be determined experimentally by propagating the same pulse at time intervals of $T_0$.
This would produce peaks at all harmonics of $\omega_0$, and their heights would be proportional to 
$|\tilde{f}(n\omega_0)|$, giving us sufficient information for computing the required ratio between any
pair of even and odd harmonics. Thus in order to determine $\delta k([2m-1]\omega)$, two measurements need to
be made which are, (1) of $R_{mn}$ done by sending alternating pulses, perturbed and unperturbed, at 
every $T_0/2$, and (2) the ratio between the peaks at the corresponding even and odd harmonics of
the unperturbed pulse done by propagating the same pulse at intervals of $T_0$.  

\section{A table-top experiment replicating the procedure}
This process may be replicated with the help of a pair of identical waveguides, some dielectric material 
that can fill up a part of one waveguide, a signal generator, a spectrum analyzer and a fast switch. 
We can generate a periodic pulse and let it propagate through one of the waveguides, or alternate 
between waveguides with and without a dielectric with the help of the fast switch. In this manner 
the two step experimental process described in the previous section may be carried out. The 
thickness of the dielectric may be altered, resulting in a change in value of $R_{mn}$. The value of
$\delta k$ may be determined using the values of $R_{mn}$ and the ratio of the peaks of even and 
odd harmonics as described in the previous section, and this may be compared with the expected value 
as predicted by the properties of the dielectric material. It is important for the electrodes that launch
the wave to be well matched with the arriving signal. In addition, the propagation and the reception of 
the signal need to be efficient enough so that the odd harmonics are well above the noise floor. 
The use of amplifiers at the input and/or output may be useful in reducing the signal to noise ratio.  
Demonstrating this concept through such an experiment would help establish confidence in applying
it to an operating storage ring. This table top experiment can be done easily, and interpretation 
of the results is straight forward as the dielectric properties remain the same for a given 
frequency through out their propagation. The procedure and analysis associated with performing
such a measurement in an operating storage ring is more involved and is the topic of the next section. 

\section{Generalization to a dynamic perturbation in the dispersion}
In a storage ring, the beam creates a cloud density that varies with time and position, leading to a
dispersion relation that is dynamic. Let us assume that the storage ring has a single train of bunches,
and all the electron cloud created by the beam is cleared away, {\it ie} none of it is trapped. We propose
launching a pulse at two instances during a revolution period. Once after the train has passed the
launch point, with a certain delay $\tau$, and half a revolution period later when the beam is
half way around the ring, when presumably there is no electron cloud along the length of propagation
of the pulse. The pulse is propagated a distance $x$ before being detected, and the detector
is shielded from the signal produced by the beam using a fast gate.  We assume that the tail of the 
last bunch in the train passes the launch point at $t=0$, and $x=0$ corresponds to the launch point. 
Rather than an abrupt appearance and disappearance of the dielectric medium as in the table-top 
experiment we have a periodic buildup, and dissipation of the dielectric medium which here is the 
electron cloud. The perturbation in the wave number is now a function of $t$ and $x$ with periodicity 
$T_0$. We denote this as $\Delta K$, with $\delta k$ as the function that repeats itself at intervals 
of $T_0$.  In this case, $\delta k$ is a function of $\omega$, $t$ and $x$ and is a consequence of 
the spatial and temporal dependence of the cloud density. We may express the relationship between 
$\Delta K$ and $\delta k$ as
\begin{equation}
 \Delta K = \sum_{n=-\infty}^{\infty} \delta k (\omega,t-nT_0-\frac{x}{v_b}) 
\end{equation}  
The term containing $v_b$, the velocity of the beam causes the medium to follow the beam. 
Let us define $t^{\prime} = t - x/v_b$. Then following the steps analogous to Eq~(1-5), we have
\begin{equation}
  \Delta K(\omega,t^{\prime}) = \sum_{n=-\infty}^{\infty} K_n(\omega){\rm e}^{in\omega_0t^{\prime}}
\end{equation}
where
\begin{equation}
 K_n(\omega) = \frac{1}{T_0} \tilde{\delta k}(\omega,n\omega_0) 
\label{Kn-exp}
\end{equation}
We may then rewrite $\Delta K$ as
\begin{equation}
  \Delta K(\omega,t,x) =  \sum_{n=-\infty}^{\infty} K_n(\omega) {\rm e}^{in\omega_0(t-x/v_b)} 
\label{DK-expn}
\end{equation}
In this section, we require that the pulse shape is symmetric about a mid point. We denote
the pulse delay $\tau$ as the time period between $t=0$ and time of generation of the mid 
point of the pulse. The signal from this delayed pulse launched between intervals of $T_0/2$ may be 
expressed as
\begin{equation}
 F(t,\tau) = \sum_{n=-\infty}^{\infty} f(t - \frac{nT_0}{2} - \tau)  
\end{equation}  
as before, this gives
\begin{equation}
 F(t,\tau) = \sum_{n=-\infty}^{\infty} F_{2n} {\rm e}^{i2n\omega_0(t-\tau)} 
\end{equation}
with $F_{2n} = (2/T_0)\tilde{f}(2n\omega_0)$. 
The propagated pulse shape, with the delay $\tau$ taken into account would be
\begin{equation}
  f(t,x,\tau) = \frac{1}{2\pi}\int_{-\infty}^{\infty} d\omega \tilde{f}(\omega) 
{\rm e}^{-i[k(\omega)x - \omega (t-\tau)]}
\end{equation}
further, the signal produced by the propagated pulses will be given by 
\begin{equation}
 F(t,x,\tau) = \sum_{n=-\infty}^{\infty} F_{2n} 
{\rm e}^{-i[k(2n\omega_0)x - 2n\omega_0(t-\tau)]}
\end{equation}

The perturbation of the pulse shape may be obtained by replacing
$\delta k$ by $\Delta K$ into Eq~(\ref{df-expr}), and using 
Eq~(\ref{DK-expn}). We then have
\begin{eqnarray}
 \delta f(x,t,\tau) = & & -\frac{ix}{2\pi}\int_{-\infty}^{\infty}d\omega \sum_{n=-\infty}^{\infty} K_n 
(\omega) {\rm e}^{in\omega_0(t-x/v_b)} \nonumber \\ 
& & \times \tilde{f}(\omega) {\rm e}^{-i[k(\omega)x-\omega t + \omega \tau]} 
\end{eqnarray} 
taking a Fourier transform of this gives
\begin{eqnarray}
& & \tilde{\delta f}(\omega,t,\tau) = -\frac{ix}{2\pi}\int_{-\infty}^{\infty}\int_{-\infty}^{\infty} 
dt d\omega^{\prime}  \nonumber \\ 
& & \times  \sum_{n=-\infty}^{\infty} K_n (\omega^{\prime}) {\rm e}^{in\omega_0(t-x/v_b)}  \nonumber \\
& &  \times \tilde{f}(\omega^{\prime}) {\rm e}^{-i[k(\omega^{\prime})x-\omega^{\prime} t + 
\omega^{\prime} \tau]} {\rm e}^{-i\omega t} 
\end{eqnarray}
Interchanging the order of summation and integration, and integrating over
$t$ by using the usual expression for the dirac-delta function, that was 
used in Eq~(\ref{Snxexp}), we can reduce this to
\begin{eqnarray}
 & & \tilde{\delta f}(\omega,t,\tau) = -ix\sum_{n=-\infty}^{\infty}\int_{-\infty}^{\infty}d\omega^{\prime} 
K_n (\omega^{\prime})\tilde{f}(\omega^{\prime}) \times \nonumber \\
& & {\rm e}^{-in\omega_0 x/v_b}{\rm e}^{-i[k(\omega^{\prime})x + \omega^{\prime} \tau]} 
\delta(n\omega_0 + \omega^{\prime} - \omega) 
\end{eqnarray} 
From Eq~(\ref{df-signal}), we know that the $n^{th}$ harmonic amplitude of the signal produced
by $\delta f$ is $(1/T_0)\tilde{\delta f}(n\omega_0,x,\tau)$. Thus, by performing the above integration
we obtain
\begin{eqnarray}
& & \delta F_n(x,\tau) = -\frac{ix}{T_0}\sum_{m=-\infty}^{\infty} K_m([m-n]\omega_0) 
 {\rm e}^{-im\omega_0x/v_b}  \nonumber \\
& & \times {\rm e}^{-i[k([m-n]\omega_0)x + (m-n)\omega_0\tau]} \tilde{f}([m-n]\omega_0)   
\end{eqnarray}
By substituting $(m-n) = N$, this may be further simplified to
\begin{eqnarray}
& & \tilde{\delta F}_n(x,\tau) =  -\frac{ix}{T_0}{\rm e}^{-in\omega_0x/v_b}
\sum_{N=-\infty}^{\infty}K_{N+n}(N\omega_0)  \nonumber \\ 
& & \times {\rm e}^{-i[k(N\omega_0)x + (\tau + x/v_b)N\omega_0]}\tilde{f}(N\omega_0)
\end{eqnarray}
The ratio between even and odd harmonics as given in Eq~(\ref{sig-ratio}) will now 
be far more complex.
Some simplification would occur because we have required $f(t)$ to be symmetric
about $t=0$. As a result of this $\tilde{f}(\omega)$ is purely real. This would
lead to $|\tilde{f}[(2m-1)\omega_0,x,\tau]| = \tilde{f}[(2m-1)\omega_0]$ as we 
know that variation in $x$ and $\tau$ only leads to a phase shift. Consequently,  
\begin{eqnarray}
 R_{mn} & & \approx \frac{|\delta F_{2m-1}(x,\tau)|}{|F_{2n}(x,\tau)|} = 
 x \left | \sum_{N=-\infty}^{\infty}K_{N+2m-1}(N\omega_0) \times \right . \nonumber \\ 
& & \left . {\rm e}^{-i[k(N\omega_0)x  +  (\tau + x/v_b)N\omega_0]}
\frac{\tilde{f}(N\omega_0)}{2\tilde{f}(2n\omega_0)} \right | 
\label{sig-ratio2}
\end{eqnarray}   
Typically, it is only the absolute values of the harmonics of $F$ that are 
measurable. Using the fact that $\tilde{f}(\omega)$ is purely real, the quantity 
$\tilde{f}(N\omega_0)/\tilde{f}(2n\omega_0)$ may be measured by propagating
pulses at every $T_0$ when the beam is half way around the ring, and all the
cloud has presumably cleared away. This would give values of $\tilde{f}(n\omega_0)$
for all $n$, even and odd. In general, the components of $K$ will have real and 
imaginary parts since $\delta k$ is need not be symmetric about any value of $t$.
Thus the Fourier transform of $\delta k$ will be complex, which implies from
Eq~(\ref{Kn-exp}) that the components of $K$ will also be complex. 

The unknown quantities in Eq~(\ref{sig-ratio2}) are the components of $K$.
To be able to determine them up to a sufficiently high order, we need to
solve a large enough set of simultaneous equations. One can generate a 
sufficient number of such equations by varying $m$, or by varying $\tau$ in steps,
and inserting the values into Eq~(\ref{sig-ratio2}), which will need to
be truncated at a large enough value in $N$. These equations are nonlinear
because the components of $K$ have real and imaginary parts.  
The set of equations can be solved using a standard numerical scheme such as the 
Newton-Raphson method. This will require initial guess values which will undergo 
an iterative process till they converge to a set of values that satisfy the
equations. The guess values may be obtained from some idea of a
typical cloud build up pattern. The cloud decay time itself can be readily
estimated by scanning over a range of $\tau$. When $\tau$ is large enough
so that the electron cloud has decayed away, we get the
signal produced by pairs of identical pulses, where the perturbed signal
seen at the odd harmonics are absent. Knowing the components of $K$ will give us 
the periodic variation of the perturbation in the dispersion, which is $\delta k$.
From this one can determine the time variation of the cloud density if we
know the relationship between $\delta k$ and the electron density. 

The dispersion relation for a waveguide filled with a dielectric medium is well
known. For an electron cloud in a region free of external magnetic fields, the
dispersion relationships have been worked out in Refs~\cite{sonnadPAC},\cite{sonnadNIM}.
It is also shown that the relationship between $\delta k$ and the electron density
is linear for frequencies that are not too close to the waveguide cutoff frequencies.
In addition, the linear relationship is valid for electron densities corresponding
to plasma frequencies that are small compared to the wave frequency and the cutoff
frequency. These approximations are reasonable for typical conditions of electron
cloud generation in accelerators. 

\section{Summary}
This paper proposes a method to perform time resolved measurements of electron cloud
density using transverse electric pulses. Given that this is a far more sensitive
measurement compared to earlier methods that used continuous waves, it offers various
challenges. Since the duration of the pulse is much smaller than its periodicity,
the average power in the transmission can be low. One needs to determine if the
loss during the transmission is significant, and if it depends on the frequency.
A frequency dependence in the rate of attenuation in the transmitted power would 
distort the measurement. A significant advantage over the measurements using 
continuous waves is that the signal produced by the particle beam can be completely 
shielded. The strong signal from the beam increases the level of the noise floor, 
and its absence would contribute toward improving the signal to noise ratio. It 
needs to be determined from experimental tests if such a measurement requires specialized 
instrumentation instead of beam position monitors (BPM) that have typically be 
in use. Additionally, a method of suppressing a possible reflection of the pulse after 
its transmission over the required distance might be necessary.  The method and the 
concept presented in this paper are novel and will prove to be a powerful diagnostic 
tool for electron clouds in particle accelerators, if the challenges mentioned above 
could be overcome.
\bibliography{TEwavePulse}
\end{document}